\documentstyle[epsf,11pt]{article}
\newcommand{\bea}{\begin{eqnarray}}
\newcommand{\eea}{\end{eqnarray}}
\begin{document}

\newcommand{\lsim}
{{\;\raise0.3ex\hbox{$<$\kern-0.75em\raise-1.1ex\hbox{$\sim$}}\;}}
\newcommand{\gsim}
{{\;\raise0.3ex\hbox{$>$\kern-0.75em\raise-1.1ex\hbox{$\sim$}}\;}}

\begin{flushright}
{\large 
{HIP-2002-56/TH}\\
{\Large \tt hep-ph/0302141}\\
}
\end{flushright}
\vspace{0.1cm}

\begin{center}
{\LARGE\bf ZZH coupling : A probe to the origin of EWSB ?}
\\[15mm]
{\bf Debajyoti Choudhury$^{a}$, Anindya Datta$^b$ and Katri
Huitu$^{b,c}$}
\\[4mm]
$^a$Harish Chandra Research Institute, Chhatnag Road, Jhusi\\
Allahabad - 211 019, India\\[4mm]
$^b$Helsinki Institute of Physics \\
P.O.Box 64, FIN-00014 University of  Helsinki, Finland \\[4mm]
$^c$High Energy Physics Division, Department of Physical Sciences,
\\
P.O.Box 64, FIN-00014 University of  Helsinki, Finland \\[7mm]
\date{}
\end{center}

\begin{abstract}

We argue that the $ZZH$ coupling constitutes a simple probe 
of the nature of the scalar sector responsible for electroweak
symmetry breaking. We demonstrate the efficacy of this measure 
through an analysis of four-dimensional models containing scalars 
in arbitrary representation of $SU(2) \times U(1)$, as well as 
extra-dimensional models with a non-factorizable geometry. A possible 
role for the $t \bar t H$ couplings is also discussed.

\end{abstract}

\vskip 7mm

\section{Introduction} 
     \label{sect:intro}
Despite its tremendous success, the Standard Model (SM) still lacks
any direct experimental information regarding the electroweak
symmetry breaking (EWSB) sector.  While the Higgs mechanism is a very
plausible candidate, giving masses, as it does, to both the weak gauge
bosons as well as the fermions, it predicts one or more physical
spin-$0$ states, none of which has been found so far.  Detection of
these scalars (pseudo-scalars) is one of the major goals of the
particle physics experiments being built at the moment.

This very lack of experimental information precludes any definitive
statement regarding the Higgs sector.  While the minimal choice of a
$SU(2)$ doublet, as in the SM, is an aesthetically pleasing one, there
is very little support for such a choice in terms of either
phenomenological evidence or any theoretical understanding. Indeed,
many alternatives to the SM have an extended Higgs sector. In
particular, supersymmetric theories, proposed to cure the bad high
energy behaviour of the Higgs mass in SM, necessarily have at least
two scalar doublets taking part in the EWSB. Somewhat similar is the situation
with grand unified theories (GUT). Apart from containing 
typically more than one Higgs responsible for EWSB, 
such scenarios often include 
extra low-mass scalars that do not directly take part in EWSB, but mix
with the Higgs bosons thereby affecting their properties.  Since
search strategies for the Higgs(es) depend crucially on the possible
decay channels, any such mixing will have very definite
consequences.  Examples are afforded by the recently proposed models
for TeV scale gravity.  The low-energy effective theory obtained upon
compactification of the extra space time dimensions typically contain
one or more scalars that do not contribute to the EWSB, but may mix
with the SM Higgs, thus modifying the Higgs decay patterns in general.

In future experiments, we may see a number of neutral scalars. This
would immediately imply that the Higgs sector is not the minimal one as
chosen in the SM.  On the other hand, nature may choose to have an
extended scalar sector, but by a conspiracy of circumstances,
experiments may find only one scalar.  The important question is
finding out what a detected scalar, or a set of scalars, can tell us
about the mechanism of symmetry breaking~\cite{BMP,KKW}.  Effects of
an extended Higgs sector were also the subject of discussion in
refs.\cite{kundu,KKW,BMP}.  In ref.\cite{KKW}, for example, the focus
was on differentiating between various supersymmetry breaking
scenarios, if the Higgs mass were determined to be around 115 GeV, the
current lower limit for the SM Higgs \cite{LEP_higgs}. The authors of
ref.\cite{BMP}, on the other hand, studied the $HZZ$ coupling in some
detail keeping in mind various scenarios for EWSB.    

It, obviously,  would be useful to construct a simple observable 
that would be sensitive to the nature of the EWSB
mechanism. To this end, we concentrate on the coupling of the
scalar(s) to the $Z$-bosons. Unlike the fermion couplings to
the Higgs, this measure (alongwith the $HWW$ couplings) receives contributions
from any non-singlet scalar vacuum
expectation value (VEV). 
Moreover, the experimental measurement of such couplings is of particular  
importance as these drive Higgs productions through of the Bjorken
process~\cite{jdb} and gauge boson fusion~\cite{jp}.

While we have seemingly dismissed the fermion couplings to the
physical scalars, they deserve a closer examination, particularly the
ones involving the third generation.  At the LHC, for example, the top
Yukawa coupling would be measured with an accuracy $\cal{O}$(20\%) for
Higgs lighter than about 130 GeV \cite{dz}.  At a linear collider
operating at $\sqrt{s}=750$ GeV and with an integrated luminosity 50
fb$ ^{-1}$, the same coupling is likely to be measured with an
accuracy of $\cal{O}$(10\%) for $m_H<240$ GeV \cite{MO}. However, even an
observed deviation from the SM value would not be enough to
distinguish a scenario with more than one doublet Higgs from an exotic
one. Some indirect inferences may be drawn though. For example, if the
measured value of the top-Yukawa at the EW scale turns out to be
significantly larger than the SM value, then, under renormalization group
evolution, it would approach a non-perturbative regime much
sooner than within the SM. This would be indicative of some 
new physics operative at
that scale. Such an eventuality is a real possibility in scenarios
wherein more than one scalar takes part in the EWSB. As the scalar(s)
giving mass to top quark must then have a VEV (VEVs) smaller than the SM Higgs
VEV, the top-Yukawa must be enhanced accordingly.  Thus the structure
of scalar sector and top Yukawa are highly correlated.

The plan of the rest of the article is as follows. In 
section \ref{sect:higgses}, we
will discuss the coupling of the Higgs to a pair of $Z$-bosons in
extensions of SM wherein the scalar sector consists of, 
apart from the usual doublet of $SU(2)_L$, fields transforming 
in higher-dimensional representations. Also discussed, in this context, 
are constraints emanating from perturbativity of the top quark-Higgs 
coupling as also from the naturalness of cancellations in the 
$\rho$-parameter.
In Section \ref{sect:extradim}, we concentrate on the
models of TeV scale gravity, especially those incorporating a 
non-factorisable geometry. We analyse the mixing of the 
graviscalars with the SM Higgs and the consequent modification
of the couplings to the $Z$. Also studied is supersymmetric variation of 
such models. Finally we conclude in section 4.

\section{$ZZH$ coupling in four dimensional models} 
    \label{sect:higgses}
On account of its excellent agreement with almost all of current
experimental data, the Standard Model serves as the natural template
for any investigation of the Higgs sector. Within the SM, the $ZZH$
coupling follows from the kinetic term $(D_\mu\phi)^\dagger
D^\mu\phi$, where $\phi$ is the Higgs doublet and $D_\mu$ denotes the
gauge covariant derivative. Clearly, if the Higgs sector were to be
extended, one needs consider a sum of such individual kinetic terms,
with the covariant derivative defined appropriately for each type of
field.

There are good reasons for studying models with several scalars.
Not only do such scalars 
occur in almost any extension of the Standard Model, even their
representations may vary considerably. Perhaps the simplest such extension
is afforded by models for spontaneous CP violation~\cite{weinberg}. Then, 
there are the models for neutrino masses (and lepton number violation) 
which incorporate $SU(2)_L$ triplets, of both the real and the complex 
kind~\cite{GM}. 
A further set of models are those that lead to gauge 
unification at high energy scales, 
either directly into a single gauge group or via some intermediate steps.
Such scenarios, perforce, have to have additional Higgses 
that break the large symmetry down to the SM gauge group. 
The details, of course, depend on the choice of the initial gauge symmetry 
as well as the chain of symmetry breaking. As an example, let us consider the 
simplest model for gauge unification, namely $SU(5)$. The SM fermions 
are then contained in two disjoint representations, viz. the 
${\bf{\bar 5}}$ and ${\bf{10}}$.
Since 
\begin{equation}
{\bf{\bar 5}}\times{\bf 10}={\bf 5}+{\bf 45},\;\; {\bf 10}\times{\bf
10}={\bf{\bar 5}}+{\bf 45}+{\bf 50}\;\;{\rm and}\;\; {\bf{\bar
5}}\times{\bf{\bar 5}}={\bf{\overline{10}}}+{\bf\overline{15}},
\end{equation}
one can have gauge invariant  {\em renormalizable} Yukawa 
couplings for scalars transforming as one of
${\bf{5}}$, ${\bf{10}}$, ${\bf{15}}$, ${\bf{45}}$, and
${\bf{50}}$. Of these, only ${\bf{\bar 5}}, {\bf{\overline{45}}}$ and 
${\bf{\overline{15}}}$ contain neutral scalars ---
the first two in the doublet of the embedded $SU(2)_L$
while the last as a $Y=1$ triplet ---
and may generate fermion or gauge boson masses.
Apart from the abovementioned set of fields, 
one must also consider a scalar in the adjoint 
representation (${\bf 24}$) as that is the one responsible for 
breaking  $SU(5)$ down to the Standard Model gauge group. 

A particular problem that a generic GUT model faces is that 
of generating the correct low-energy mass spectrum. To facilitate 
this, in a natural manner, it is often necessary to consider 
Higgses in other representations. While, at the tree level,
such scalars would couple to only the gauge bosons and the other 
(lower-dimensional) scalars in the theory:
\begin{equation}
{\bf{\bar 5}}\times{\bf 45}={\bf 24}+\dots,\;\;
{\bf{\overline{15}}}\times{\bf 15}={\bf{70'}}+\dots,\;\;
{\bf{\overline{45}}}\times{\bf 45}={\bf 35}+\dots \ ,
	\label{other_reps}
\end{equation}
these may also participate in non-renormalizable couplings involving 
fermions. It can be easily ascertained that these new scalars
appearing in eq.(\ref{other_reps}) contain, in addition to the 
SM gauge group singlet, components transforming as
one of triplets (hypercharge $Y = 0, 1$), 4-plets ($Y = 3/2, 1/2$) 
or 5-plets ($Y= 2,1,0$) of $SU(2)_L \otimes U(1)_Y$
\footnote{The intermediate mass scalars, which couple to the gauge
 bosons, will change the running of the gauge couplings.
 E.g. the effect of new representations on $\sin^2\theta_W$
 has been considered in \cite{rg_higgs}.  The measured value
 of the weak mixing angle restricts the mass scales and
 representations of the new scalars.}.

A third set of examples,
where more than one physical scalar emerges, is afforded by the
supersymmetric models.
If the Standard Model (SM) description were to be valid up to high energy
scales, it would face the theoretical problem 
of the Higgs boson mass receiving potentially large 
radiative corrections.
To counter this in nonsupersymmetric GUTs, one has to precisely fine tune
between the bare mass and radiative corrections.
Supersymmetry, on the other hand, offers a very natural
solution to the problem, and without coming into conflict 
with any experimental observation.
A major difference between the Standard Model and supersymmetric
models is that the Higgs sector is necessarily extended
to include at least two Higgs doublets instead
of just the one needed in the Standard Model.

Having argued that it is natural for extensions of the Standard Model 
to contain more than one Higgs with a neutral component, let us now 
consider the Higgs couplings with the $Z$. Summing over all possible 
neutral Higgses, we have
\begin{equation}
L_{Z-H}= (g^2+g'^2) Z^\mu Z_\mu \sum_i 
	\left[ \frac 12 T_{3i}^2 H^0_i H^0_i  +
         T_{3i}^2  v_i H^0_i  +
         \frac 12 T_{3i}^2 v_i^2 \right] \ ,
	\label{L_zh}
\end{equation}
where $T_i$'s are the isospins of the Higgs multiplets 
and $v_i$'s are the VEVs of the corresponding neutral components.
Since the last term in (\ref{L_zh}) gives the $Z$ mass,
\begin{equation}
	m_Z^2 = (g^2+g'^2) \sum_i T_{3 i}^2  v_i^2 \ ,
\label{Zmass} 
\end{equation}
it is obvious that, in the presence of many Higgses with nonzero VEVs,
the magnitude of an individual VEV must be smaller than that 
of the SM one. Similar conclusions follow from the mass of the $W$-boson 
as well. In fact, one of the strongest constraints on VEVs of non-standard
Higgses arises from the precision measurement of the 
electroweak $\rho$ parameter. Defined as the ratio of the strengths of the 
charged-- and the neutral currents, we have, at the tree level,
\[
	\rho \equiv \frac{m_W^2}{m_Z^2 \cos^2 \theta_W} \ ,
\]
with a Standard Model value of unity. 
While quantum corrections to the $\rho$-parameter have been calculated, 
for our analysis it suffices to consider only the tree level expression.
In the presence of many Higgses,  we have
\begin{equation} \displaystyle
	\rho = {\sum_i r_i \left[ T_i (T_i + 1) - T_{3 i}^2 \right] v_i^2 
		 \over
		\sum_i 2 T_{3 i}^2 v_i^2 
	       }.
	\label{eq:rho}
\end{equation}
with $r_i = 1/2$ for real representations and 1 otherwise.
It is immediately apparent that the experimental observation of 
$\rho \approx 1$ can be naturally satisfied only for certain very 
specific choices of the Higgs representations. For all others, a degree 
of fine-tuning between the VEVs is necessary. Such a fine-tuning 
may be motivated though from other considerations~\cite{GM}. 
Throughout the rest of the paper,
we will assume that the $\rho$-parameter should be
unity at the tree-level, thereby defining a constraint 
in the space of Higgs VEVs.

We now revert back to the 
coupling of the $Z$ with the generic CP-even neutral Higgs $H^0_i$. These 
may be conveniently parametrized by 
\begin{equation}
	L_{H_i ZZ} =  c_i \; \frac{g m_Z}{\cos\theta_W} \;H^0_i Z_\mu Z^\mu \ ,
          \label{defn:c_i}
\end{equation}
where $c_i$ is a measure of the scaling with respect to the 
SM Higgs coupling. A sum rule for the $c_i$s follows:
\begin{equation}
	{\cal C} \equiv \sum_j c_j^2 = 4 {\sum_i T_{3 i}^4 v_i^2 
		 \over
		\sum_i T_{3 i}^2 v_i^2 
	       } \ .
	\label{eq:coup_sum}
\end{equation}
Clearly, $\sum_j c_j^2 $ increases if representations with larger 
hypercharge get relatively larger VEVs.
On the other hand, the presence of an arbitrary number of doublets 
(and singlets) does not change ${\cal C}$ from its SM value of unity.
In fact, ${\cal C}$ is bounded from both above and 
below\footnote{Singlet VEVs, of course, are not being considered here.}
\begin{equation}
	4 \: {\rm max} (T_{3i}^2) \geq {\cal C} \geq 
	4 \: {\rm min} (T_{3i}^2) 
	\label{C_limits}
\end{equation}
with the extremal values being reached in the event of a particular 
VEV dominating all others. A clear departure from the Standard Model 
relation ${\cal C} = 1$ would thus indicate the presence of new physics.

Were we interested only in maximizing $\left| {\cal C} - 1 \right|$ while 
maintaining $\rho = 1$, the result would be given by eq.(\ref{C_limits}). 
However, apart from the consequent fine-tuning amongst the VEVs of the 
higher representations, a further aspect needs to be considered. 
As we have already remarked,  eq. (\ref{Zmass}) implies that 
were higher representation to have nonzero VEVs, 
the magnitude of doublet VEV(s) would necessarily 
be smaller than the Standard Model value. An immediate consequence is that 
the Yukawa couplings leading to the fermion mass terms would need 
to be scaled up. 
Restricting  ourselves to the simplest (and {\em conservative}) case
of only one Higgs doublet ($\Phi_{1/2}$) 
being responsible for the fermion mass generation, 
the Yukawa couplings are now given by 
\begin{equation}
	y_f = y_f^{\rm SM} {v_{\rm SM} \over v_{1/2} } \ .
	\label{Yukawa_scaling}
\end{equation}
Future experiments, both at the LHC \cite{dz} and at a
linear collider \cite{nlc} will serve to measure the
Yukawa couplings for the top- and bottom-quarks as well as the
$\tau$-lepton, at least for intermediate values of the Higgs
mass\footnote{Note that the analyses of these papers implicitly
    assume that the
    Higgs couplings, whether gauge or Yukawa, are relatively close
    to the SM ones. A breakdown of this assumption could lead to
    significant changes in the results obtained.}.
{\em Assuming that the theory contains only one doublet},
this information could then be used to set stringent limits on parameters
such as $v_{1/2} / v_{\rm SM}$.  In other words, such measurements 
would immediately alert us about the presence of a nonminimal 
Higgs sector.
In fact, an accurate measurement of more than
one Yukawa coupling could also shed light on the presence of more than
one doublet. We will later consider the explicit example of a triplet
Higgs model.

Is the analysis of this paper redundant then? Note that if the Yukawa
couplings are found to be very close to the SM ones, it is extremely
unlikely that a renormalizable four-dimensional theory could
accommodate a value of ${\cal C}$ significantly different from
unity. Identification of a second neutral scalar and a measurement of
an anomalous value of ${\cal C}$ would then immediately point towards
either a radion or a SM singlet with higher-dimensional couplings with
the SM fields. On the other hand, if the Yukawas are found to be quite
different from their SM values, the only unambiguous conclusion that
we may draw from such measurements is the presence of {\em some}
nontrivial structure of the EWSB sector. An analysis as proposed in
this paper could, once again, prove very useful in our quest to
unravel the mystery of this sector.

\begin{figure}[!h]
\vspace*{-15em}
\centerline{\hspace*{3em}
\epsfxsize=13cm\epsfysize=15.0cm
                     \epsfbox{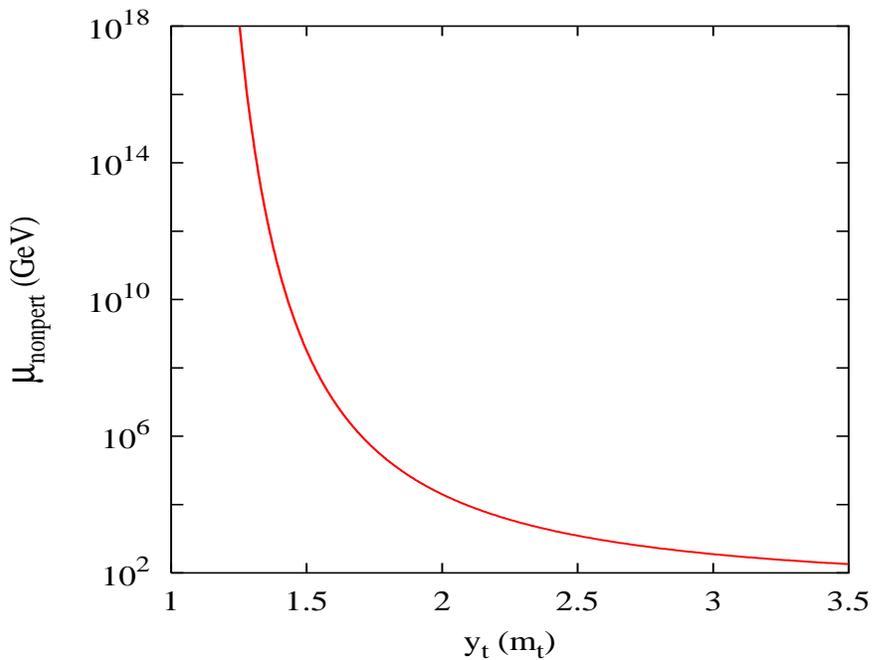}
}
\caption[]{\em The scale at which the top-quark Yukawa coupling becomes 
	nonperturbative (see eq.(\protect\ref{eq:nonpert})) as a function of 
	its value at the pole.}
\label{fig:nonpert}
\end{figure}

Concentrating on the top Yukawa, a larger value at low energies would
imply a faster growth as one evolves it to larger energies. 
The one-loop\footnote{While higher loop results are available, it 
	is sufficient, for our purposes, to consider only the 
	lowest order expression.}
renormalization group equation for the top Yukawa reads
\begin{equation}
	2 \pi {{\rm d}\: y_t^2 \over {\rm d}\: \ln \mu}
		= - \left[8 \alpha_3 + {9 \over 4} \alpha_2 + 
			{17 \over 12} \alpha_1 \right] y_t^2
		+ {9 \over 8 \pi} y_t^4 \ ,
\end{equation}
where $\mu$ is the scale and $\alpha_i$ represent the (scale dependent) 
gauge couplings within the Standard Model. 
Clearly, if $y_t (m_t)$ were to be larger, the $y_t^4$ piece on 
the right hand side would tend to dominate, thereby 
driving up the Yukawa coupling as the scale increases. The 
scale of nonperturbativity, $\mu_{\rm nonpert}$, may be defined through
\begin{equation}
	y_t(\mu_{\rm nonpert}) = \sqrt{4 \pi} \ .
	\label{eq:nonpert}
\end{equation}
In fig.\ref{fig:nonpert}, we exhibit the functional dependence of 
$\mu_{\rm nonpert}$ on the value of the top-Yukawa at the
pole, namely $y_t(m_t)$. As the figure clearly shows, if we require 
the theory to be perturbative until very high scales, $y_t(m_t)$ 
would be required to assume values close to its Standard Model magnitude. 

Having developed the tools, let us now proceed to investigate, numerically, 
the extent to which the sum ${\cal C}$ may vary in the presence of 
extra scalars, while satisfying experimental bounds. Rather than complicate 
matters by including scalars in arbitrary representations, we restrict 
ourselves to scenarios wherein the Standard Model doublet is accompanied 
by either $SU(2)_L$ triplets or fourplets or fiveplets. Within each such 
class, we do not further restrict the value of the hypercharge $Y$. We 
could, of course, have considered extra doublets (or singlets), but such
inclusion would not have virtually any bearing on our numerical results.

\vspace*{1cm}

\noindent
\underline{\it Triplet Higgses.}

\vspace*{3ex}
\noindent
Apart from the doublet [(2, 1/2)], we now have
an arbitrary number of real [(3, 0)] and complex [(3, 1)] triplets. 
Clearly, a VEV for the former increases $\rho$ 
(see eq.(\ref{eq:rho})) from its Standard Model value of unity, 
while that for the latter serves to reduce it.
Thus, experimental consistency stipulates that 
the presence of a non-zero VEV for one kind must be accompanied 
by a VEV for at least one of the other kind. 

\begin{figure}[t]
\vspace*{-15em}
\centerline{\hspace*{3em}
\epsfxsize=11cm\epsfysize=14.0cm
                     \epsfbox{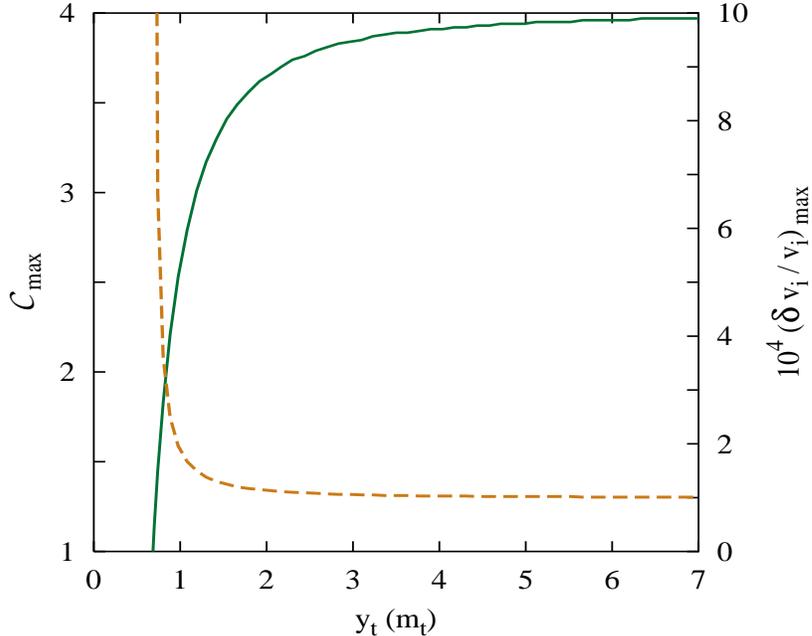}
}
\caption[]{\em The solid line represents the
		quantity ${\cal C}$ as a function of the top Yukawa 
           for a model with the standard (one doublet) Higgs sector 
	   augmented by one each of real and complex triplets. 
	   The tree level constraint $\rho = 1$ has been imposed. 
	   For an arbitrary number of triplets satisfying this constraint,
	   ${\cal C}$ is restricted to be {\em below} this curve.
	   The dashed line gives the maximum fractional deviation 
	   of a triplet Higgs VEV that is allowed by the data 
	   at 95\% C.L.}
\label{fig:triplet}
\end{figure}

For the sum of the squared couplings, we now have $ 1 \leq {\cal C}
\leq 4 $ with the upper limit being reached when at least one of the
complex triplets (and hence at least one of the real triplets) have a
VEV much larger than the doublet.  An examination of
eq. (\ref{eq:coup_sum}) shows that the presence of a multitude of
triplets would not change the upper bound on ${\cal C}$ in any way.
Rather, for a given scaling of the top Yukawa (i.e. for the VEV of the
doublet), the maximal value of ${\cal C}$ is reached when the extra
contribution to the gauge boson masses accrues from only a single set
of triplets.  In our quest of identifying the maximal possible
deviation from the value of ${\cal C}$, we may thus limit ourselves to
a study of just one such pair. The results are presented in
fig. \ref{fig:triplet} as a function $y_t(m_t)$. The width to the
curve allowed by the experimental uncertainties in the measurement of
$\rho$ is too small to be discernible.  As expected, for $y_t(m_t)
\approx y_t^{\rm SM}(m_t)$ (namely a very small scaling of the doublet
VEV), ${\cal C} \approx 1$. On the other hand, even for an arbitrarily
large $y_t(m_t)$, which corresponds to the bulk of the gauge boson
masses arising from the triplet VEVs, ${\cal C}$ only reaches its
asymptotic value of 4.

Until now, we have entirely desisted from remarking on the fine-tuning
problem in the cancellation of the two (or more) triplet contributions
to the $\rho$-parameter.  The degree of fine-tuning may be quantified
in terms of the maximal fractional deviation, $(\delta v_i / v_i)_{\rm
max}$, that a given triplet VEV $v_i$ may suffer without coming in
conflict with the experimental results. In fig. \ref{fig:triplet}, we
also show the `bounds' corresponding to a 95\% C.L. agreement with the
data~\cite{PDG}.  As is immediately apparent, if we wish the
fine-tuning to be no worse than {\em per mille} level, we would be
confined to $y_t (m_t) \lsim 0.72$ and hence to ${\cal C} \lsim 1.3$,
a value quite close to the SM one.  However, if we admit a fine-tuning
of $\sim 10^{-4}$, the whole range for ${\cal C}$ would be admissible.

A constraint that we have not yet imposed is the requirement of
perturbativity. As fig.~\ref{fig:nonpert} amply demonstrates, $y_t
(m_t) \lsim 3.5$ for perturbation theory to make sense even at the
top-mass scale. This still allows ${\cal C} \lsim 3.5$. On the other
hand, if we demand that the theory remain perturbative until close to
the GUT scale, we are immediately restricted to ${\cal C} \lsim
2.5$. This still represents a very significant deviation from the
Standard Model.

Before delving into the discussion of more complex Higgs sectors, we
will briefly mention how the experimentally measured Higgs widths 
can constrain the Higgs sector. 
An illustrative example is the triplet Higgs model by Georgi and
Machachek (GM) \cite{GM}.  Couplings of the SM-like Higgs of the model
to gauge bosons and fermions are suppressed/enhanced with respect
to the SM case by a factor of $v_{1/2} / v_{\rm SM}$.
Therefore, measurement of these factors
at future experiments can shed light on the physical model.
E.g. the partial decay width into 
a fermion pair can be expressed in terms of the SM width as
$\Gamma_{GM}^{ff} = (v_{\rm SM} / v_{1/2})^2 \;\Gamma_{SM}^{ff}$.
Assuming the ratio of the VEVs is not too different from unity, 
the signature profile at a collider, and hence the experimental efficiencies, 
would not change drastically. 
One can use the experimentally measured value ($\Gamma^{exp}_{ff}$) of
the $ff$ width and its error ($\Delta \Gamma^{exp}$) to constrain
the ratio of VEVs at a certain confidence level using the inequality
\bea 
\frac{\Gamma^{exp}_{ff} - \alpha \Delta \Gamma^{exp}}
{\Gamma^{SM}_{ff}} \lsim ( v_{\rm SM} /v_{1/2})^2\lsim 
\frac{\Gamma^{exp}_{ff} + \alpha \Delta \Gamma^{exp}}
{\Gamma^{SM}_{ff}},
\label{contr}
\eea
where  $\alpha$ depends on the confidence level at which one would 
want to constrain
the ratio. 

The projected accuracy on $\Gamma (h \rightarrow \tau \tau)$
measurement at the LHC is of the order of $10 - 20 \%$ \cite{dz}. 
An $e^+ e^-$ linear collider fares much better in this respect. The
accuracy \cite{nlc} of the Higgs width measurement can be as good as
5\% instead.
From the above expression, eq. (\ref{contr}),
it is evident how the experimental accuracy determines the constraint on
the ratio.

\vspace*{0.7cm}

As the results for 4-plet Higgses are qualitatively very similar to
those obtained above, we move straightaway to the 5-plets.

\vspace*{0.5cm}
\begin{figure}[!h]
\vspace*{-15em}
\centerline{\hspace*{3em}
\epsfxsize=11cm\epsfysize=14.0cm
                     \epsfbox{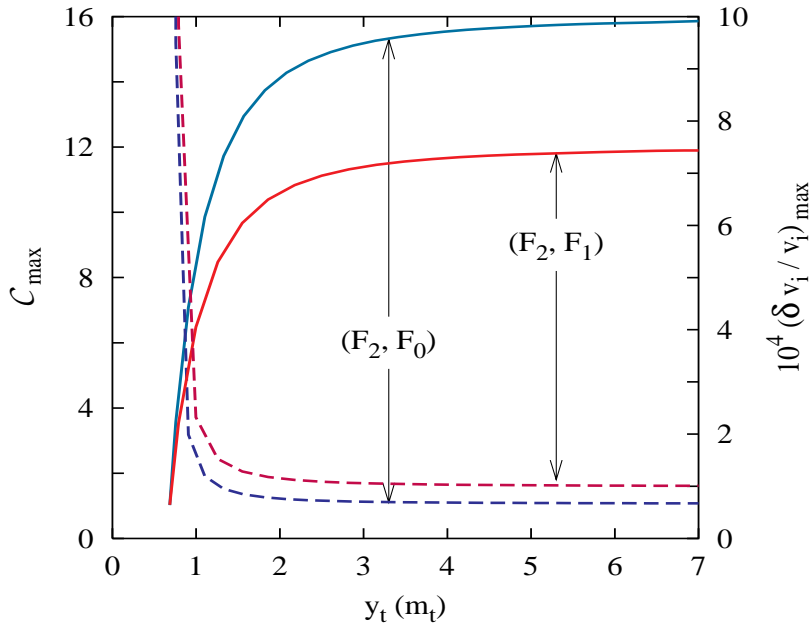}
}
\caption[]{\em The quantity ${\cal C}$ (solid lines) 
	   as a function of the top Yukawa 
           for a model with a doublet Higgs and arbitrary number of 
	   real and complex 5-plets and 
	   respecting the tree level constraint $\rho = 1$. 
	   The two curves 
	   correspond to the cases with no $F_0$ type VEV and no $F_1$ 
	   type respectively. 
	    The dashed lines give the maximum fractional deviation 
	   of a 5-plet Higgs VEV that is allowed by the data 
	   at 95\% C.L.
	   If both kinds are present, the 
	   corresponding curves would lie in between the two respective sets.}
\label{fig:5plet}
\end{figure}

\noindent
\underline{\it 5-plet Higgses.}

\vspace*{3ex}
\noindent

Apart from the doublet [(2, 1/2)], we now have an arbitrary number of
real [$F_0 \equiv (5, 0)$] and complex [both $F_1 \equiv (5, 1)$ and
$F_2 \equiv (5, 2)$] 5-plets. Clearly, VEVs for the first two types
can only enhance $\rho$ while that for a $F_2$-type field suppresses
it. Thus, to satisfy the experimental constraint in presence of 5-plet
VEVs, one needs at least one VEV of the (5, 2) kind.

For the sum of the coupling squares, we now have the 
inequality
\begin{equation}
1 \leq {\cal C} \leq 16 \ ,
\end{equation}
with the upper limit being reached when at least one of the $F_2$'s
(and hence at least one of the other triplets) have a VEV much larger
than the doublet. Once again, for a given $y_t(m_t)$, the maximal
value of ${\cal C}$ is reached when a single $F_2$-type field acquires
a large VEV and, for simplicity, we shall illustrate our results under
this assumption. There are two interesting limits though. If no $F_0$
gets a VEV, ${\cal C}(y_t)$ follows a particular limit
\[ 
{\cal C}(F_2, F_1) 
	 = \frac{288 \langle (5, 2) \rangle^2 + \langle (2, 1/2) \rangle^2}
                {24 \langle (5, 2) \rangle^2 + \langle (2, 1/2) \rangle^2} \ ,
\]
with an ensuing maximum of 12 (see fig.\ref{fig:5plet}).
This is irrespective 
of the actual numbers of $F_2$'s and $F_1$'s, as along as at least 
one of each kind exists. In the opposite 
case (no $F_1$ gets a VEV), ${\cal C}$ moves along the other extreme 
\[ 
{\cal C}(F_2, F_0) 
         = \frac{256 \langle (5, 2) \rangle^2 + \langle (2, 1/2) \rangle^2}
                {16 \langle (5, 2) \rangle^2 + \langle (2, 1/2) \rangle^2} \ ,
\]
and reaches the absolute maximum of 16. 
Both these curves are displayed in fig.~\ref{fig:5plet}. 
The results for the mixed case lie somewhere in between these two 
extremes depending on the relative magnitudes of the VEVs.
Also displayed in the same figure are the fine-tuning curves corresponding
to either of these branches. Once again, if we limit ourselves to a fine-tuning
no worse than $10^{-3}$, we are limited to ${\cal C} \lsim 2$. 
 
The arguments concerning perturbativity 
remain essentially the same as in the case of triplets.

\section{$ZZH$ coupling in higher dimensional models}
      \label{sect:extradim}
Theories with TeV-scale gravity \cite{ADD,RS} have recently been
proposed as an alternative to supersymmetry as the solution for the
hierarchy problem. Such models, perforce, have to be defined in
dimensions larger than four. Since the results of the present day
gravity experiments demonstrate excellent agreement with Newtonian
gravity down to a distance scale about 0.1 mm, the compactification
radius for the extra dimensions must be smaller than this scale. In
the simplest examples of such models, all the SM fields are assumed to
be confined to 4-dimensional hypersurfaces or branes.

Two different explanations for the perceived hierarchy between the
Planck-scale and the fundamental ($\sim$ TeV) scale exist.  In the
first class of models~\cite{ADD}, the metric in the higher dimensional
theory is assumed to be a factorizable one, and the large volume of
the compactified dimensions leads to the aforementioned
hierarchy. Such models are characterized by very closely spaced
Kaluza-Klein tower(s) of the graviton.

The model proposed by Randall and Sundrum (RS)~\cite{RS}, on the other
hand, is based on the premise that the 5-dimensional theory is
described by ${\rm AdS}_5$ geometry. With the 4-dimensional part of
the metric suffering an exponential warp depending on the fifth
co-ordinate, the perceived scale of gravity on two branes separated in
the fifth direction would naturally differ enormously. The lowest
Kaluza-Klein excitation of the graviton is now at the TeV
scale. Interestingly, the model also contains an extra scalar, the
{\em radion}, which may be lighter than the graviton excitations and
may also mix with the SM Higgs thereby changing the relevant
phenomenology. It is this class of models that we shall concentrate on
in this section.

While the models with TeV scale gravity were originally deemed as
alternatives to low energy supersymmetry, it was subsequently realized
that the two ideas could well be linked to each other. For one, the
very stabilization of the brane-world picture may need underlying
supersymmetry, especially if they are to represent vacua of
superstring theories.  At a more phenomenogical level, theories with
extra dimensions seem to provide a natural way of communicating
supersymmetry breaking from a hidden sector onto our
world~\cite{LS,CL}.

We will, hence, consider both the non-supersymmetric and
supersymmetric versions of the RS-model.

\vspace*{1cm}

\noindent
\subsection{Non-supersymmetric RS-model.}
      \label{sect:nonsusyRS}

\vspace*{3ex}

\noindent
In the model proposed in ref. \cite{RS}, possibly the lowest mass
excitation from the gravitational sector is a scalar, the radion. This
corresponds to the 4-dimensional manifestation of $g_{55}$ of the
full 5-dimensional metric and  can also be visualised as the distance
between the two branes situated at the two fixed points of the
orbifold on which the extra dimension is compactified.  
As is obvious, the radion needs to be 
stabilized for the effective 4-dimensional Planck mass to {\em remain}
exactly what we perceive it to be. 
In Ref.~\cite{GW}, Goldberger and Wise demonstrated that the inclusion 
of a bulk scalar, coupling minimally, to gravity can lead to an effective
potential for the radion and hence to its stabilization. The mass of the 
radion turns out to be lower 
than that for the lowest $J=2$ Kaluza-Klein excitation,
and hence the radion is likely to offer the first signature 
of the brane world picture.

It can be shown from very general considerations that 
the radion couples to the
SM fields via the trace of the energy momentum tensor.  
In addition, the effective action may contain terms leading to 
curvature-Higgs mixing \cite{GRW,CGK}. 
At the two-derivative level, the coupling of the gravity 
and the Higgs $H$ --- confined to the visible brane --- may be 
parametrized as 
\begin{equation}
S=-\xi \int d^4 x \sqrt{-g_{vis}} \: R(g_{vis}) \: H^\dagger H,
\end{equation}
where the Ricci scalar $R(g_{vis})$ is obtained from the  induced 
four dimensional metric on the visible brane. Clearly, the above term 
induces a kinetic mixing between the Higgs and
the radion.  Additional mixing is introduced by the fact that both 
the neutral component of the Higgs ($h$) as well as the radion ($\phi$) 
acquire VEVs:
\begin{equation}
	\langle H \rangle = v, \quad 
	\langle \phi \rangle = \Lambda, \quad {\rm with} \quad
	\gamma \equiv \frac{v}{\Lambda} \ .
\end{equation}
To obtain fields with canonical quantization rules, it is necessary 
to effect field redefinitions:
\begin{equation}
\pmatrix{ \phi \cr H \cr} \to \pmatrix{ \phi' \cr H' \cr} 
	\equiv Z_R \: {\cal M}^{-1} \pmatrix{ \phi \cr H \cr}  \ ,
\qquad
{\cal M} = \pmatrix{ \cos \theta  & -\sin \theta \cr
                     Z_R \sin \theta-6\xi \gamma \cos \theta 
                                 & Z_R \cos \theta +6 \xi \gamma \sin \theta
		    \cr}
\end{equation}
where 
\begin{equation} \displaystyle
Z_R^2=1-6\xi \gamma^2 (1+6\xi) \quad {\rm and} \quad 
\tan 2 \theta = \frac{12 \xi \gamma Z_R  m_h^2}{m_H^2(Z_R^2-36\xi^2\gamma^2)-
m_{\phi}^2}.
\end{equation}

In the limit $\xi \rightarrow 0$, we recover back the SM Higgs from
$H^{'}$.  Requiring that the kinetic terms for the physical fields
$H'$ and  $\phi'$ be positive,  restricts us to~\cite{CGK}
\begin{equation}
\frac{-1}{12} \left( 1+\sqrt{1+4/\gamma^2} \right) \: \le \xi \: \le
\frac{1}{12} \left(\sqrt{1+4/\gamma^2}-1 \right).
\label{limits}
\end{equation}
For $\Lambda=1$ TeV, this translates to $-0.75 <\xi <0.59$, while for
$\Lambda=10$ TeV, the constraint is much weaker ($-6.7 <\xi < 6.6$).

The couplings of the physical scalars to a generic gauge boson $V$ 
can be expressed as~\cite{CHKY}
\begin{equation}
{\cal L}=-  \: \frac{M_V^2}{v}
\: V_{\mu} V^{\mu} \: 
 \left[ \left( {\cal M}_{22} + \gamma {\cal M}_{12} \right)
		 \: H
             + \left( \gamma {\cal M}_{11} +  {\cal M}_{12} \right)
		 \: \phi
	\right] \ .
\label{Ldiag}
\end{equation}
The parenthetical coefficients directly give the strength of the
corresponding interaction when compared to the case with no
mixing\footnote{Note that we neglect here the small correction due to
the conformal anomaly.}, and hence correspond to the quantities
$c_{H'}$ and $c_{\phi'}$ (see eq.~(\ref{defn:c_i})).  An important
observation needs to be made here. Even in the absence of any
Higgs-curvature mixing ($\xi = 0$), the radion couples to the gauge
bosons {\em without} affecting their coupling to the Higgs.  Since the
latter coupling is responsible for unitarizing the gauge boson
scattering amplitudes ($V_i V_j \to V_k V_l$) and since the
radion-exchange contribution to these amplitudes are quite similar to
the Higgs-exchange ones, it is obvious that the presence of the radion
would destroy the partial wave unitarity for such
amplitudes~\cite{HKM,CCGM}.  This is quite unlike the case of the
multi-Higgs models in renormalizable 4-dimensional
theories. Quantitatively, the unitarity constraint has only a weak
dependence on $\gamma$~\cite{HKM} and can be approximated by
$|\xi|\lsim 2.7$.

\begin{figure}[h]
\begin{center}
\mbox{\epsfxsize=9.truecm\epsfysize=9.truecm\epsffile{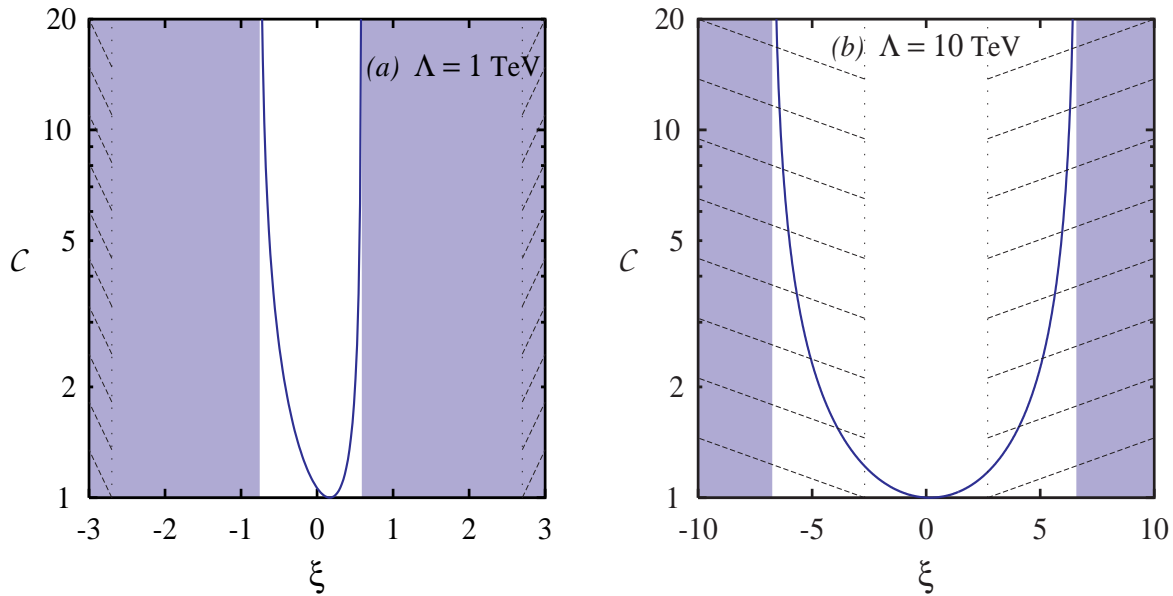}}
\end{center}
	\vspace*{-10ex}
\caption{\em The parameter ${\cal C}$ as a function of radion-Higgs
mixing parameter $\xi$ for two different values of $\Lambda$, the
radion VEV.  The shaded area in each plot is ruled out by the
requirement that the kinetic energy be positive definite, while the
hatched area would lead to a loss of partial wave unitarity.}
	\label{crad}  
\end{figure}

Reverting back to our measure for distinguishability, we
now have
\begin{equation} \displaystyle 
{\cal C}  =  
	\left( {\cal M}_{22} + \gamma {\cal M}_{12} \right)^2
             + \left( \gamma {\cal M}_{11} +  {\cal M}_{12} \right)^2
    = 1 + \frac{\gamma^2}{Z_R^2} \left( 1 - 6 \xi \right)^2 \ ,
\end{equation}
which clearly indicates a deviation from the SM result of ${\cal C} =
1$ {\em even in the absence of curvature-Higgs mixing}. However, the
extent of deviation, in such cases, is very small and is unlikely to
be detectable.  The presence of a mixing term, though, can enhance
the effect manifold, as is borne out by fig.~\ref{crad}. That the
effect should decrease with increasing $\Lambda$, the radion VEV, is
expected. The figure also amply demonstrates the relative importances
of the constraints from the positivity of the kinetic energy and from
perturbative unitarity.  At this stage, it should be noted that the
radion-gauge boson coupling is the driving force behind a large ${\cal
C}$. In other words, it is the $\phi' Z Z$ coupling ($\phi'$ is the
radion-dominated scalar) that grows very fast with increasing $\xi$
whereas the $H'ZZ$ coupling never grows beyond its SM counterpart.

\begin{figure}[!h]
\vspace*{-10em}
\centerline{\hspace*{3em}
\epsfxsize=10cm\epsfysize=13.0cm
                     \epsfbox{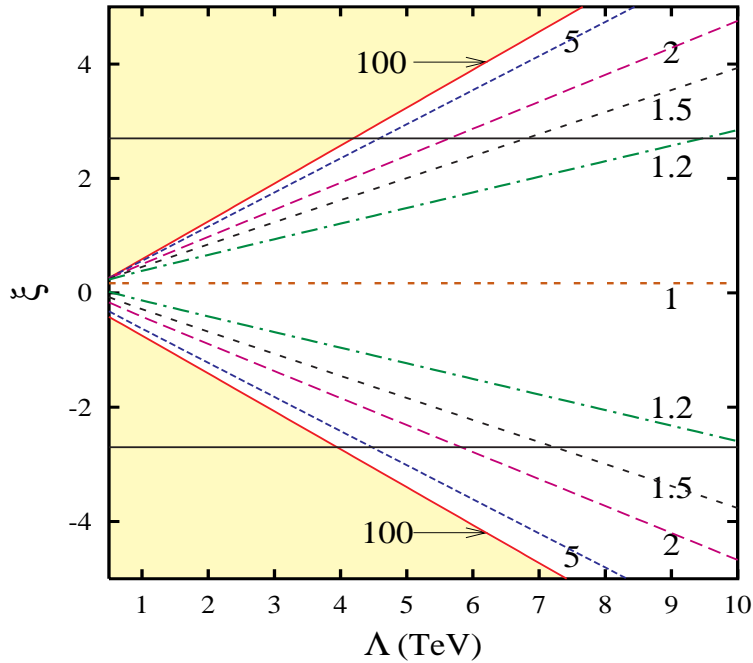}
}
\caption[]{\em Contours of constant ${\cal C}$ in the 
	$\Lambda$-$\xi$ plane for the nonsupersymmetric RS scenario. 
	The shaded regions correspond to the unphysical domain of 
	negative kinetic energy. Requiring that gauge-boson scattering 
	amplitudes respect perturbative unitarity would restrict one 
	to the region {\em between} the two horizontal solid lines.}
\label{fig:cont}
\end{figure}

From an experimental point of view, it is worth asking how much of the
parameter space one could possibly distinguish from the Standard
Model.  The answer, of course, would be a function of the sensitivity
of the experiment under consideration. An indication, though, may be
obtained from fig.~\ref{fig:cont}, wherein we have plotted iso-${\cal
C}$ contours in the $\Lambda$-$\xi$ plane. The contours are almost
linear, with deviations from linearity becoming apparent only for
$\Lambda \sim v$.  The contour corresponding to absolute
indistinguishability from the SM (${\cal C} = 1$) is but the straight
line $\xi = 1/6$.  As one moves away from ${\cal C} = 1$, the contour
splits up into two disjoint curves. For large ${\cal C}$, the curves
asymptotically reach the boundary separating the unphysical domain ($Z
= 0$).  It is quite apparent that our measure (${\cal C}$) is unlikely
to differentiate a RS theory with large $\Lambda$ unless it is also
accompanied by a large Higgs-curvature mixing. For radion VEVs of a
few TeV, on the other hand, marked deviations from the SM expectation
would be obtainable even for relatively small mixings.

\subsection{Supersymmetric RS-model.}
           \label{sect:susyRS}
\vspace*{3ex}
\noindent
A particularly important issue for Randall-Sundrum-type models relates to 
the (phenomenologically needed) stabilization of the radion.
Several schemes have been proposed, especially in the 
context of the supersymmetric versions~\cite{FLP,LS,CL}.
Interestingly, in such proposals, supersymmetry breaking and radius
stabilization are intimately linked to each other.

As a concrete example of such scenarios, we 
shall consider in this work the low energy model discussed in
ref. \cite{CEN}.
The standard SUGRA particle content needs to be supplemented by a
complex scalar 
field\footnote{Here $\langle r \rangle$ is the radius
    of compact $S^1$ and $\langle b \rangle$ the Aharanov-Bohm phase of
    the graviphoton around the $S^1$.} 
$T=k\pi (r +ib \sqrt{2/3})$~\cite{LS}.  
In addition, one needs matter in the bulk in order to ensure
supersymmetry breaking~\cite{FLP}.  This matter is represented by
a universal hypermultiplet $S$.  In the model of ref.\cite{CEN}, the warp
factor is given by 
\begin{equation}
e^{-\langle T+T^\dagger \rangle}=\frac{\Lambda^2}{3M_P^2},\\
\end{equation}
where $\Lambda={\cal{O}}$(TeV) and $T=\langle T \rangle+t/\Lambda $.
Here $t$ is the radion superfield, which will mix with the Higgs
fields through the kinetic and mass terms.  The term in the
Lagrangian relevant to the matter chiral superfields on the 
visible brane is given by
\begin{equation}
{\cal L} =\frac 12 G_{ij}D_\mu \phi_i D^\mu \phi_j,\nonumber
\label{scalL}
\end{equation}
where $G_{ij}$ is obtained from the effective K\"ahler potential 
in the usual way:
\begin{equation}
\begin{array}{rcl}
G_{ij}& = & \displaystyle \frac{\partial^2 K_{eff}}{\partial
\phi_i\partial\phi_j} \ ,\\[2ex]
K_{eff}& = & \displaystyle  \Lambda^2 \exp{\left\{-\frac{t+t^*}{\Lambda
}+\frac{1}{\Lambda^2}\Sigma_{eff} (\{\phi_i\})\right\} } \ .
\end{array}
    \label{kahler}
\end{equation}
with $\Sigma_{eff} (\{\phi_i\} ) =\Sigma_i |\phi_i|^2 +\lambda (H_1\cdot
H_2 + h.c. )$.

Since the field $t$ is a gauge singlet, the kinetic term does not 
lead to a $tZZ$ coupling. However, such a term does arise from 
the trace of the energy-momentum tensor and is given by
\begin{equation}
L_{tZZ}=-tZ_\mu Z^\mu \frac{m_Z^2}{\Lambda}.
\end{equation}
For the MSSM scalars $H_1$ and $H_2$, on the other hand, we now have
\begin{equation}
L_{HZZ}= \left( 1+(\lambda+1)(\lambda +3)\frac{v^2}{4\Lambda^2}\right)
\frac{g}{4\cos\theta_W}(v_1 H_1^0 + v_2 H_2^0)Z_\mu Z^\mu,
\end{equation}
where $v_i$ are the VEVs of $H_i^0$ and $v^2=v_1^2+v_2^2$.
The mass of the $Z$ boson is then
\begin{equation}
m_Z=\frac {g v}{ 2 \cos\theta_W} \:  \left[
1+(\lambda+1)(\lambda +3)\frac{v^2}{4\Lambda^2}\right]^{1/2} \ .
\end{equation}
An interesting feature of this model relates to the ratio of the 
VEVs for $H_i^0$, namely $\tan\beta \equiv v_2/v_1$. The potential 
is minimized for $\tan\beta =1$ \cite{CEN}. This is quite unlike 
the case of the MSSM wherein $\tan \beta$ is a free parameter, 
and current experimental data strongly disfavour $\tan\beta  \sim 1$.
This difference can be traced back to the additional contributions
that the Higgs mass receives from its coupling to the radion (vide 
eq.\ref{kahler}). 
A consequence of $\tan \beta = 1$ is that, of the two 
orthogonal combinations
\begin{equation}
\phi^0=\frac {1}{\sqrt{2}} (H_1^0 +H_2^0)  \quad {\rm and} \quad 
     H^0=\frac {1}{\sqrt{2}} (H_1^0 -H_2^0) \ , 
\label{scalars}
\end{equation}
$\phi^0$ alone contributes to the $Z$ mass, while the 
coupling of the $H^0$ to $Z$ vanishes identically. 
Note that the redefinition of eq.(\ref{scalars}) does not 
entirely diagonalize the kinetic terms. To achieve that goal, one has 
to effect a further field redefinition~\cite{CEN} akin to that 
performed in Section \ref{sect:nonsusyRS} and involving all three 
of $\{H_1^0,H_2^0,t\}$. On doing this, we obtain, for the sum 
of the squares of the scalar couplings to the $Z$ (and normalized
to the SM value), 
\begin{equation} \displaystyle
{\cal C}= 
    \left[ 1 + \frac{v^2}{2 \Lambda^2} (1+\lambda) 
      \left\{ \frac{1}{\sqrt{2}} - \: \frac{\lambda +5}{8} 
      \right\} \right] ^2 
   + 
     \frac{v^2}{\Lambda^2} \: 
     \left[ 1 + \frac{(1+\lambda)}{2\sqrt{2} }
       -\frac{v^2}{16\Lambda^2}(1+\lambda)(1-3\lambda )\right]^2 \ .
\label{eqsusyrs}
\end{equation}

\begin{figure}[h]
\centerline{\hspace*{3em}
\epsfxsize=9cm\epsfysize= 9cm
                     \epsfbox{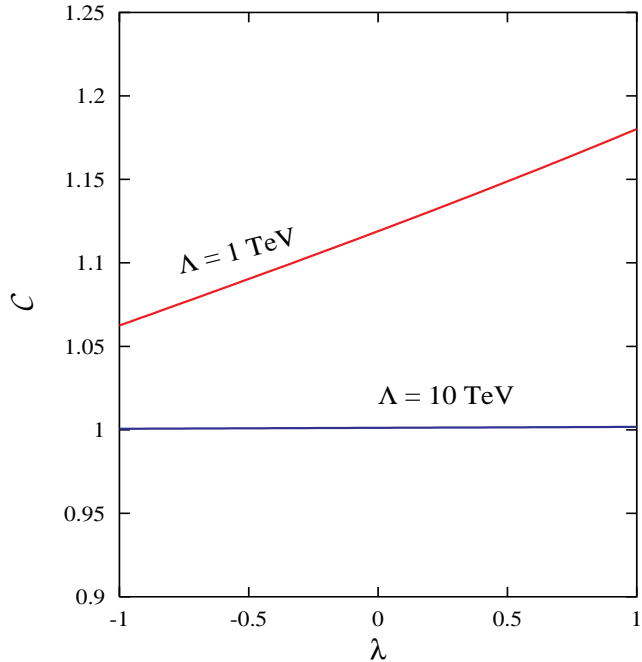}
}
\caption{\em The parameter ${\cal C}$ as a function of 
parameter $\lambda$. 
The two lines correspond to $\Lambda =1$ TeV and $\Lambda =10$ TeV.
respectively}
\label{crad3}  
\end{figure}

In fig.\ref{crad3}, we present ${\cal C}$ as a function of the Higgs 
mixing parameter $\lambda$. It is evident from eq. (\ref{eqsusyrs}), 
that $\cal C$ grows with $\lambda$.  One sees easily that, for a 
large $\Lambda$, the deviation from the Standard Model value of unity
is negligible. However, for a relatively small $\Lambda$, the effect
might be noticeable. 

\section{Discussion and conclusions}

To summarize, we have studied the $ZZH$ coupling in models with
different types of scalar sectors, but with the Standard Model gauge
group. We establish that the sum of the squares of the couplings of the 
scalars to the $Z$ can be an efficient discriminator. For scenarios with 
non-zero VEVs for Higgses in higher dimensional representations of $SU(2)_L$,
this quantity can differ substantially from the SM value of unity. 
However, if we demand either naturalness in the value for the $\rho$-parameter
or demand that the top Yukawa coupling stays perturbative until GUT-scale,
this parameter is restricted to values close to one. 

In higher dimensional models with warped geometry, on the other 
hand, the effect of
Higgs-curvature mixing on this parameter may be quite significant. This 
remains true even after the constraints from perturbative unitarity 
are imposed. Thus, such a model would be clearly distinguishable 
from a ``natural'' multi-Higgs model provided the radion VEV is less 
than a few TeVs. Imposition of supersymmetry on such models, however, 
renders this measure largely ineffective. In either case, 
the radion, being a gauge singlet,
does not contribute to fermion masses and consequently 
no constraints emerge from this sector.

Before we conclude, we would like to make some brief comments about the $ZZH$
couplings in the recently proposed model of `the littlest Higgs' 
\cite{littlehiggs}. The local gauge symmetry, at high energies, is 
enlarged to $\left[ SU(2) \otimes U(1) \right]^2$. 
This is broken down to the familiar $SU(2)_L \otimes U(1)_Y$
at a scale $\Lambda \sim {\cal O}$(TeV).
The scalar potential at low energy is
generated by mechanism very similar to that of Coleman and Weinberg
\cite{CW} and the details have been 
presented in ref.\cite{han_little}. The physical spectrum includes
two neutral CP even, one neutral CP odd, one charged and one
doubly-charged scalars.
In contrast to the RS or SUSY-RS models, the extra scalars are not
gauge singlets any more and all the fermions, in general, can have
extra contributions to their masses of ${\cal O}(v /\Lambda)$.  If we
concentrate on the couplings of the two neutral CP-even Higgses to the
pair of $Z$s, and calculate the quantity $\cal C$ as we have
done earlier, the expression looks qualitatively very similar to
eq. (\ref{eqsusyrs}).  In other words, as we increase the scale
$\Lambda$, the coupling of one of these states to a pair of $Z$ tends to
that of the SM case while the other decouples from SM. The results 
are similar to those for the SUSY-RS case. 

\begin{flushleft} {\bf Acknowledgements} \end{flushleft}

\noindent
AD and KH thank the Academy of Finland
(project number 48787) for financial support.
DC would like to thank the Helsinki Institute of Physics for hospitality 
during the period the project was initiated and the 
Department of Science and Technology, Government of 
India for financial assistance under Swarnajayanti Fellowship.

\end{document}